\documentclass[sigconf]{acmart}

\usepackage{amsmath}
\usepackage{threeparttable}
\usepackage{multirow}
\usepackage{graphicx}

\newtheorem{Def}{Definition}

\AtBeginDocument{%
  \providecommand\BibTeX{{%
    \normalfont B\kern-0.5em{\scshape i\kern-0.25em b}\kern-0.8em\TeX}}}

\copyrightyear{2021}
\acmYear{2021}
\setcopyright{acmlicensed}
\acmConference[SIGIR '21]{Proceedings of the 44th International ACM SIGIR Conference on Research and Development in Information Retrieval}{July 11--15, 2021}{Virtual Event, Canada}
\acmBooktitle{Proceedings of the 44th International ACM SIGIR Conference on Research and Development in Information Retrieval (SIGIR '21)}
\acmPrice{15.00}
\acmDOI{10.1145/3404835.3463052}
\acmISBN{978-1-4503-8037-9/21/07}

\begin{document}
\fancyhead{}

\title{Inductive Representation Learning in Temporal Networks via Mining Neighborhood and Community Influences}

\author{Meng Liu}
\affiliation{%
  \institution{Heilongjiang University}
  \city{Harbin}
  \country{China}
}
\email{2191438@s.hlju.edu.cn}

\author{Yong Liu}
\authornote{Corresponding Author.}
\affiliation{%
  \institution{Heilongjiang University}
  \city{Harbin}
  \country{China}
}
\email{liuyong123456@hlju.edu.cn}



\begin{abstract}
Network representation learning aims to generate an embedding for each node in a network, which facilitates downstream machine learning tasks such as node classification and link prediction. Current work mainly focuses on transductive network representation learning, i.e. generating fixed node embeddings, which is not suitable for real-world applications. Therefore, we propose a new inductive network representation learning method called MNCI by \textbf{m}ining \textbf{n}eighborhood and \textbf{c}ommunity \textbf{i}nfluences in temporal networks. We propose an aggregator function that integrates neighborhood influence with community influence to generate node embeddings at any time. We conduct extensive experiments on several real-world datasets and compare MNCI with several state-of-the-art baseline methods on various tasks, including node classification and network visualization. The experimental results show that MNCI achieves better performance than baselines.
\end{abstract}

\ccsdesc[500]{Information systems~Data mining}
\ccsdesc[500]{Information systems~Social networks}
\ccsdesc[500]{Computing methodologies~Artificial intelligence}
\ccsdesc[500]{Computing methodologies~Machine learning}

\keywords{Inductive Representation Learning; Temporal Networks; Neighborhood and Community Influences}


\maketitle
\section{Introduction}
\label{introduction}
In the real world, network data is ubiquitous such as social network, e-commerce network, and citation network, etc. By analyzing these network data, researchers can obtain user behavior to enable effective information retrieval. As a popular field, network representation learning (NRL) aims to represent a network by mapping nodes to a low-dimensional space \cite{cui2019a}. The node embeddings generated by NRL can be used for downstream machine learning tasks such as node classification, link prediciton and social search. 

\textbf{Related work.} Based on the training goal, we can divide NRL into transductive learning and inductive learning \cite{trivedi2019dyrep}. In the early stages of NRL, most of the methods for generating node embeddings are \textbf{\emph{transductive}}, which generate fixed node embeddings by directly optimizing the final state of the network. For example, DeepWalk \cite{perozzi2014deepwalk} performs a random walk procedure over the network to learn node embeddings, node2vec \cite{grover2016node2vec} proposes a biased random walk procedure to balance the breadth-first and depth-first search strategy, and HTNE \cite{zuo2018embedding}  learns node embeddings by using the Hawkes process to capture historical neighbors' influence.

However, although transductive learning methods have good results in downstream tasks, they have difficulty adapting to the dynamic network. Many real-world tasks require node embeddings to be updated alongside network changes. Therefore, transductive learning will have to retrain the whole network to obtain new node embeddings, which is not feasible for real-world networks, especially large-scale networks.

Unlike transductive learning, \textbf{\emph{inductive}} learning attempts to learn a model that can dynamically update node embeddings over time. For example, GraphSAGE \cite{hamilton2017inductive} learns a function to generate embeddings by aggregating features from a node’s local neighborhood, and DyREP \cite{trivedi2019dyrep} proposes a two-time scale deep temporal point process which captures the interleaved dynamics of the observed processes for modeling node embeddings. 

\textbf{Our contributions.} We propose a novel inductive representation learning method called MNCI to learn node embeddings in temporal networks. In temporal networks, the edges are annotated by sequential interactive events between nodes. In real-world, many networks contain interaction time between nodes, such as the bitcoin trading network, citation network, etc. MNCI can effectively capture network changes to obtain node embeddings at any time, by \textbf{m}ining \textbf{n}eighborhood and \textbf{c}ommunity \textbf{i}nfluences.

We believe that nodes in the network are influenced by both neighborhood and communitiy. For \textbf{\emph{neighborhood}} influence, it is obvious that historical neighbors of nodes will influence their future interactions. We will model the neighborhood influence from both neighbor characteristcs and time information.

For \textbf{\emph{community}} influence, we define several communities and learn an embedding for each community. Given a node, it may have different closeness to different communities. The deeper closeness the node is to a community, the more influence this community has on the node. For example, users on Twitter are influenced differently by different topics depending on their interests. Consumers have different preferences for different products.

Finally, we devise a new GRU-based \cite{s1997long, cho2014learning} aggregation function to integrate neighborhood influence with community influence.

We evaluate MNCI on mutiple real-world datasets and compare with several state-of-the-art baselines. The results demonstrate that MNCI can achieve better performance than baselines, which illustrates the capacity of MNCI in capturing network changes. We summarize our main contributions as follows.

(1) We propose a novel inductive representation learning method MNCI to learn node embeddings in temporal networks.

(2)We use the positional encoding technology to initialize node embedding, which can speed up the convergence speed in training.

(3) We model the neighborhood and community influences and modify the GRU framework to aggregate them.

(4) We empirically evaluate MNCI for multiple tasks on several real-world datasets and show its superior performance.

The source code and data can be downloaded from https://github. com/MGitHubL/MNCI.

\section{Method}

\subsection{Network Definition}

According to the time information of node interaction, we can formally define the temporal network.

\begin{Def}
\textbf{Temporal Network.} When two nodes interact, it will always be accompanied by a clear timestamp. A temporal network can be defined as a graph $G=(V,\ E,\ T)$, where $V$ and $E$ denote the set of nodes and edges respectively, and $T$ denotes the set of interactions. Given an edge $e(u,v)$ between node $u$ and $v$, there is at least one interaction matching $e(u,v)$, i.e.,  $T(u,v)=\{(u,v,t_1),\cdots, (u,v,t_n)\}$.
\end{Def}

We believe that two nodes may interact multiple times, and these interactions can be ordered by timestamp. When two nodes interact, we call them neighbors. The historical neighbor sequence of a node can be defined as follows.

\begin{Def}
\textbf{Historical Neighbor Sequence.} Given a node $u$, we can obtain its historical neighbor sequence $H_u$, which stores the historical interactions of $u$ up to the current moment, i.e., $H_u = \{ (v_1,t_1), (v_2,t_2),\cdots,(v_n, t_n) \}$. Each tuple in the sequence represents an event, i.e., node $v_i$ interacts with $u$ at time $t_i$.
\end{Def}

\subsection{Node Embedding Initialization}

For most network representation learning (NRL) methods, node embeddings need to be initialized before training. Unlike random initialization of these methods, we propose a \textbf{time positional encoding} method to generate node embeddings by using time information, which can speed up the convergence speed of training process.

To the best of our knowledge, the idea of positional encoding \cite{vaswani2017attention} is first proposed in Natural Language Processing (NLP) field. Considering that in many real-world scenarios, most nodes have no clear feature information for researchers to obtain prior knowledge. In this case, the initial time when node $u$ joins a network will be very useful for $u$, which should be further exploited. 

According to the initial time order, we can obtain an ordered node sequence $S_{node} = \{ u_1, u_2, \cdots, u_n\}$. Then, we use sine and cosine functions with different frequencies to define the encoding on each dimension in the node embeddings.
\begin{equation}
  \begin{split}
  PE_{(u,2i)} = \sin(u/10000^{2i/d})    \\
  PE_{(u,2i+1)} = \cos(u/10000^{2i/d})
  \end{split}
\label{split}
\end{equation}

Where $u$ is the $u^{th}$ node position number in $S_{node}$, $d$ is the dimension size of node embedding, $2i$ is the $(2i)^{th}$ dimension in node embedding, and $PE_{(u,2i)}$ is the encoding for the $(2i)^{th}$ dimension of the $u^{th}$ node embeding in $S_{node}$. Each dimension of the positional encoding corresponds to a sinusoid, and the wavelengths form a geometric progression from $2\pi$ to $10000 \cdot 2 \pi$. Let $PE_{u+k}$ and $PE_u$ be the embeddings for the $(u+k)^{th}$ node and the $u^{th}$ node in $S_{node}$ respectively. 
For any fixed offset $k$, $PE_{u+k}$ can be represented as a linear function of $PE_u$, which means that the function in Eq. (\ref{split}) can capture the relative time positions of nodes. 
In this way, we can obtain the initial node embedding $z_u^{t_0}$ of node $u$ at the initial time $t_0$ as follows.
\begin{equation}
z_u^{t_0} = PE_u = [PE_{(u,0)}, \cdots, PE_{(u, d-1)}]
\end{equation}

After obtaining the initial node embedding, we can mine neighborhood and community influences. Note that both influences of a node are calculated every time it interacts with other nodes, thus we omit the time superscript by default in the following unless we want to distinguish two variables with different timestamps.

\subsection{Neighborhood Influence}

We believe that after an interaction occurs between node $u$ and $v$, node $v$ will influence the future interactions of node $u$ with other nodes, and $u$ will also influence $v$. Given a node $u$, we assume that the influence on $u$ is not only related to neighbor's own characteristics, but also related to their interaction time. Therefore, to mine the neighborhood influence on each node, we will analyze its neighbors' embedding and interaction time, respectively.

\textbf{Affinity Weight.} We assume that there is an affinity between any two nodes, which reflects the closeness of their relationship. Given a node $u$ and its neighbor sequence $H_u$, we can calculate $u$'s affinity to different neighbors. After normalizing these affinities, the affinity weight $a_{(u,i)}$ for neighbor $i$ on node $u$ can be calculated as follows.
\begin{equation} 
a_{(u,i)} = \frac {\sigma (- \left \|z_u - z_i \right \|^2)}{\sum_{i' \in H_u} \sigma (-\left \|z_u - z_{i'}\right \|^2)}
\label{affinity}
\end{equation}

Where $\sigma$ is the sigmoid function, $H_u$ is node $u$'s historical neighbor sequence. We use negative squared Euclidean distance to measure the affinity between two embeddings. 

\textbf{Temporal Embedding.} In temporal networks, network structure and node behavior will evolve over time. Thus, learning temporal information is an important way to capture the evolutionary process of neighborhood influence. In this part, we learn a temporal embedding for two interactive nodes based on their interaction timestamp. Given an interaction ${(u, i, t_i)}$, the temporal embedding $z_{(u,i)}^{t_i}$ between two interactive nodes at time $t_i$ can be calculated as follows.
\begin{equation} \label{time1}
z_{(u,i)}^{t_i} = F(t_c-t_i)
\end{equation}

Where $t_c$ is the current time, $F(t)$ is the encoding function. For $F(t)$, we adopt random Fourier features to encode time which may approach any positive definite kernels according to the Bochner's theorem \cite{bochner1934a, mehran2019time2vec, wang2021inductive, xu2019self-attention, xu2020inductive}. 
\begin{equation} \label{time2}
F(t)=[cos( \omega_1 t), sin(\omega_1 t),\cdots,cos( \omega_{d/2} t), sin(\omega_{d/2} t)]
\end{equation}

Where $\omega = {\{ \omega_1,\cdots,\omega_{d/2}\}}$ is a set of learnable parameters to ensure that the dimension size of temporal embedding and node embedding are the same as $d$.

\textbf{Neighborhood influence embedding.} Combining affinity wei-ght and temporal embedding, the neighborhood influence embedding $NE_u^{t_n}$ of $u$ at time $t_n$ can be calculated as follows.

\begin{equation} \label{NE_u}
NE_u^{t_n} =\delta_u^{NE} \sum_{i \in H_u} a_{(u,i)}z_{(u,i)}^{t_i} \odot z_i^{t_{n-1}}
\end{equation}

Where $\delta_u^{NE}$ is a learnable parameter that regulates $u$'s neighborhood influence embedding, $z_i^{t_{n-1}}$ is the embedding of $u$'s neighbor $i$ at time $t_{n-1}$, $\odot$ denotes element-wise multiplication. To calculate the influence embedding of the current timestamp, we need to use the node embedding of the previous timestamp, which will be introduced later.

\subsection{Community Influence}

A community in a network is a group of nodes, within which nodes are densely connected.  But nodes in different  communities are sparsely linked. In real-world networks, nodes in the same community tend to have similar behavior patterns.

In this paper, we define $K$ communities $C=\{c_1, ..., c_k\}$ and learn an embedding for each community, where $K$ is a hyperparameter. Given a node $u$, it may have different affinity to these communities. The deeper affinity $u$ is to a community $c_k$, the more influence $c_k$ has on $u$. Let $z_{c_k}$ be the community embedding of community $c_k$.
For node $u$, we calculate its affinity to all communities. Then we normalize these affinities to obtain the weights of different communities' influence on $u$. In this case, a community $c_k$'s affinity weight $ a_{(u, c_k)} $ on $u$ can be calculated as follows.
\begin{equation}
a_{(u, c_k)}= \frac {\sigma (- \left \| z_u - z_{c_k} \right \|^2)}{\sum_{c_{k'} \in C} \sigma (- \left \| z_u - z_{c_{k'}} \right \|^2)}
\label{community}
\end{equation}

If a communtiy $c_k$ has the the highest affinity weight to node $u$ at time $t_n$, after updating $u$'s embedding from $z_u^{t_{n-1}}$ to $z_u^{t_n}$, we will dynamically update $c_k$'s embedding as shown in Eq. (\ref{z_{c_k}}), i.e., we consider that $u$ belongs to $c_k$ at time $t_n$. In this way, we can obtain the community to which each node belongs at any time.
\begin{equation}
z_{c_k} := z_{c_k} - z_u^{t_{n-1}} + z_u^{t_n}
\label{z_{c_k}}
\end{equation}

Finally, the community influence embedding $CO_u^{t_n}$ of node $u$ at time $t_n$ can be calculated in Eq. (\ref{CO_u}), where $\delta_u^{CO}$ is a learnable parameter that regulates $u$'s community influence embedding.

\begin{equation} 
CO_u^{t_n} = \delta_u^{CO} \sum_{c_k \in C} a_{(u, c_k)}z_{c_k}
\label{CO_u}
\end{equation}

\subsection{Aggregator Function}

The GRU network \cite{cho2014learning} can capture the temporal patterns of sequential data by controlling the aggregation degree of different information and determining the proportion of historical information to be reversed. In this paper, we extend GRU to devise an aggregator function, which combines neighborhood and community influences with the node embeddings at the previous timestamp to generate the node embeddings at the current timestamp. The aggregator function used in this paper is defined as follows.
\begin{equation}
UG_u^{t_n} = \sigma(W_{UG}[z_u^{t_{n-1}} \oplus NE_u^{t_n} \oplus CO_u^{t_n}]+b_{UG})
\end{equation}
\begin{equation}
NG_u^{t_n} = \sigma(W_{NG}[z_u^{t_{n-1}} \oplus NE_u^{t_n} \oplus CO_u^{t_n}]+b_{NG})
\end{equation}
\begin{equation}
CG_u^{t_n} = \sigma(W_{CG}[z_u^{t_{n-1}} \oplus NE_u^{t_n} \oplus CO_u^{t_n}]+b_{CG})
\end{equation}
\begin{equation}
\tilde{z}_u^{t_{n}} = \tanh(W_z[z_u^{t_{n-1}} \oplus (NG_u^{t_n} \odot NE_u^{t_n}) \oplus (CG_u^{t_n} \odot CO_u^{t_n})] + b_z)
\end{equation}
\begin{equation}
\label{gru}
z_u^{t_n} = (1-UG_u^{t_n}) \odot z_u^{t_{n-1}} + UG_u^{t_n} \odot \tilde{z}_u^{t_{n}}
\end{equation}

Where $\sigma$ is the sigmoid function, $\oplus$ denotes concatenation operator, $\odot$ denotes element-wise multiplication.  $NE_u^{t_n}$, $CO_u^{t_n}$ and $z_u^{t_n}$ are neighborhood influence embedding, community influence embedding and node $u$'s embedding at time $t_n$, respectively. $W_{UG}, W_{NG}, \\ W_{CG},W_z \in \mathbb{R}^{d \times 3d}$, $b_{UG}, b_{NG}$, $ b_{CG}, b_z \in \mathbb{R}^d$ are learnable parameters, $UG_u^{t_n}, NG_u^{t_n}, CG_u^{t_n} \in \mathbb{R}^d$ are called update gate, neighborhood reset gate, and community reset gate, respectively.

In this paper, we divide the reset gate in GRU into two reset gates, i.e., neighborhood reset gate $NG_u^{t_n}$ and community reset gate $CG_u^{t_n}$. We use $NG_u^{t_n}$ and $CG_u^{t_n}$ to control the reservation degree of neighborhood and community influence embeddings, respectively. Then, we aggregate the node embedding at the previous timestamp with reserved neighborhood and community influence embeddings to obtain a new hidden state $\tilde{z}_u^{t_{n}}$ at the current timestamp. Finally, we use $UG_u^{t_n}$ to control the reservation degree of historical information. Based on the node embedding $z_u^{t_{n-1}}$ at the pervious timestamp and the new hidden state $\tilde{z}_u^{t_{n}}$ at the current timestamp, we can obtain a node embedding $z_u^{t_n}$ at the current timestamp. In this way, we can calculate node embeddings inductively.

\subsection{Model Optimization}

To learn node embeddings in a fully unsupervised setting, we apply a network-based loss function which is shown in Eq. (\ref{loss function}) and (\ref{loss}), and optimize it with the Adam method \cite{kingma2015adam}. The loss function encourages nearby nodes to have similar embeddings while enforcing that the embeddings of disparate nodes are highly distinct. We use negative squared Euclidean distance to measure the similarity between two embeddings.
The loss function also encourages each node to have high affinity with the community it belongs to.

\begin{equation} 
L = \sum_{u \in V} \sum_{v \in H_u} L(u,v) + \sum_{u \in V} \max_{c_k \in C} (\log a_{(u, c_k)})
\label{loss function} 
\end{equation}
\begin{equation} 
L(u,v) =\log \sigma \left( -\left \|z_u^{t_n} - z_v^{t_n}\right \|^2 \right) - Q \cdot E_{{v_n} \sim P_n(v)} \log \sigma \left( -\left \| z_u^{t_n}- z_{v_n}^{t_n}\right \|^2 \right)
\label{loss}
\end{equation}

Due to the enormous computation cost, we use negative sampling \cite{mikolov2013distributed} to optimize the loss function as shown in Eq. (\ref{loss}), where $P_n(v)$ is a negative sampling distribution,  $Q$ is the number of negative samples. In Eq. ({\ref{loss function}}), we also introduce $\log a_{(u, c_k)}$ to maximize the affinity between a node and the community it belongs to.

\begin{table*}[htbp]
\centering
\caption{Node classification of all methods on all datasets}
\label{classification}
\begin{threeparttable}
\begin{tabular}{c c c c c c c c c}
\toprule[2pt]
Metric&	method&	DBLP& BITotc & BITalpha&	ML1M& AMms& Yelp\\
\midrule[1pt]
\multirow{5}{*}{Accuracy}
&	DeepWalk&	0.6140& 0.5907& 0.7294&	0.6029& 0.5780& 0.5067\\
~
& node2vec&	0.6249& 0.5958& 0.7495&	\textbf{0.6196}& 0.5772& 0.5135\\
~
& GraphSAGE&	0.6331& 0.6003& 0.7389&	0.6124& 0.5763& 0.5184\\
~
& HTNE&	0.6347& 0.5999& 0.7635& 0.5890& 0.5767& 0.5273\\
~
& DyREP&	0.6259& 0.6100& 0.7430& 0.6023& 0.5755& 0.5209\\
~
& MNCI&	\textbf{0.6395}& \textbf{0.6256}& \textbf{0.7842}&0.6137& \textbf{0.5874}& \textbf{0.5334}\\
\midrule[1pt]
\multirow{5}{*}{Weighted-F1}
&	DeepWalk&	0.6107& 0.5120& 0.6761& 0.5863& 0.4252& 0.3981\\
~
& node2vec&	0.6210& 0.5123& 0.6832&	0.5836& 0.4248& 0.4184\\
~
& GraphSAGE&	0.6239& 0.5105& 0.6750&	0.5766& 0.4216& 0.4065\\
~
& HTNE&	0.6307& 0.5109& 0.6806&	0.5415& 0.4255& 0.4180\\
~
& DyREP&	0.6203&	0.5114& 0.6843& 0.5729& 0.4248& 0.4093\\
~
& MNCI&	\textbf{0.6412}&	\textbf{0.5172}& \textbf{0.6859}&	\textbf{0.6026}& \textbf{0.4268}& \textbf{0.4293}\\
\bottomrule[2pt]
\end{tabular}
\end{threeparttable}
\end{table*}

\section{Experiments}

\subsection{Experimental Setup}

\subsubsection{Datasets}

We conduct experiments on six real-world datasets. \emph{DBLP} is a co-authorship network of computer science \cite{zuo2018embedding}. \emph{BITotc} and \emph{BITalpha} are two datasets from two bitcoin trading platforms \cite{kumar2016edge, kumar2018rev2}. \emph{ML1M} is a movie rating dataset \cite{li2020time}. \emph{AMms} is a magazine rating dataset \cite{ni2019justifying}. \emph{Yelp} is a Yelp Challenge dataset \cite{zuo2018embedding}.

\subsubsection{Baselines}

We compare MNCI with five state-of-the-art baseline methods, i.e., Deepwalk, node2vec, GraphSAGE, HTNE and DyREP. The details of these methods are described in Section \ref{introduction}.

\subsubsection{Parameter Settings}

We set the embedding dimension size $d$, the learning rate, the batch size, the number of negative samples $Q$, and the community number $K$ to be 128, 0.001, 128, 10, and 10 respectively, and other parameters are default values.

\subsection{Results and Discussion}

\subsubsection{Node Classification}

We train a Logistic Regression function as the classifier to perform 5-fold cross-validation that predicts node labels, and use Accuracy and Weighted-F1 as metrics. As shown in Table \ref{classification}, MNCI outperforms all other baselines over six datasets. Compared with GraphSAGE and HTNE that use neighborhood interactions, MNCI focuses on both neighborhood and community influence, leading to further performance improvements.

\begin{figure}[htbp]
\centering
\begin{minipage}[t]{0.16\textwidth}
\includegraphics[width=1\textwidth]{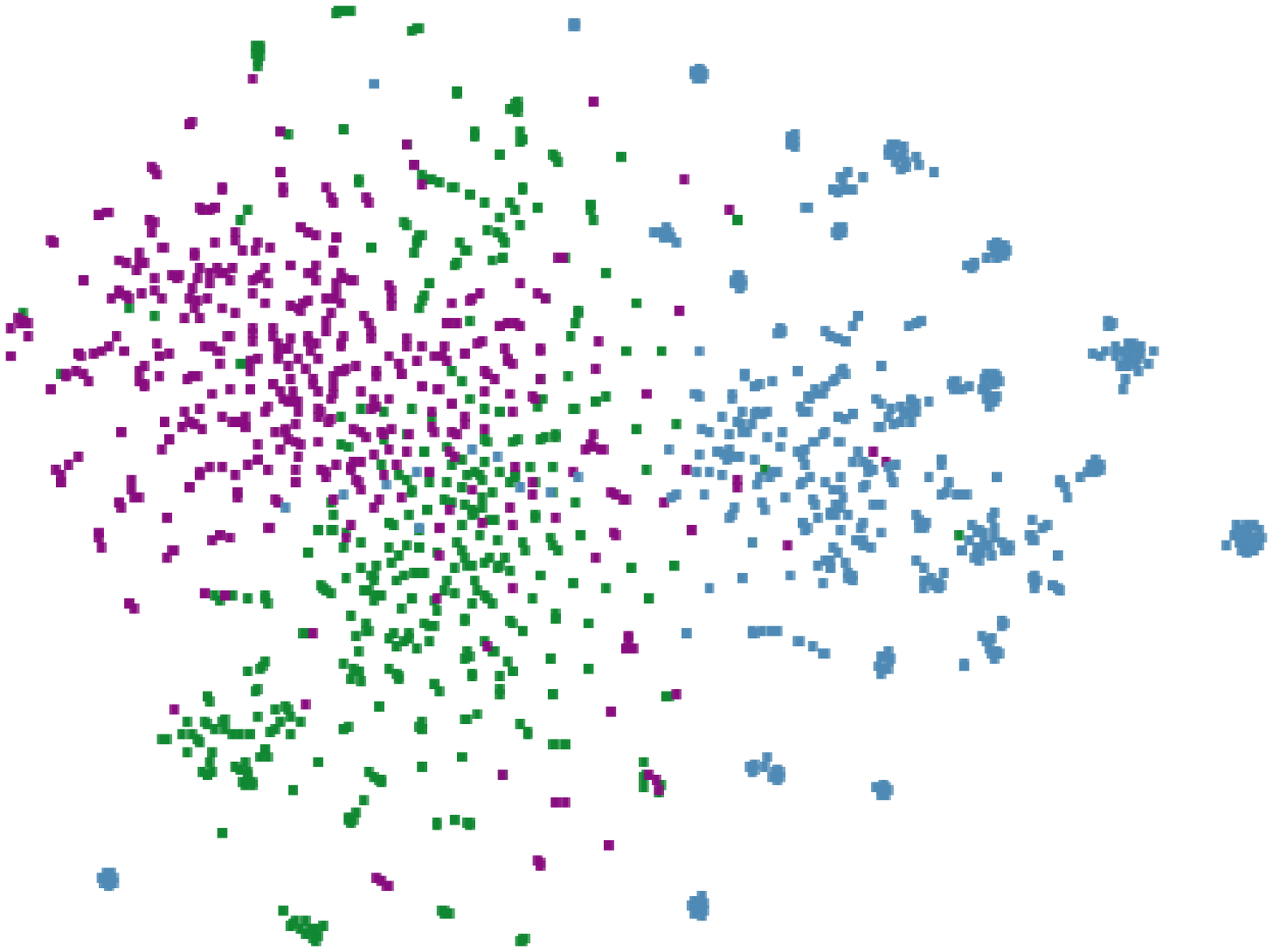}
\centerline{(a) DeepWalk}
\end{minipage}%
\begin{minipage}[t]{0.16\textwidth}
\includegraphics[width=1\textwidth]{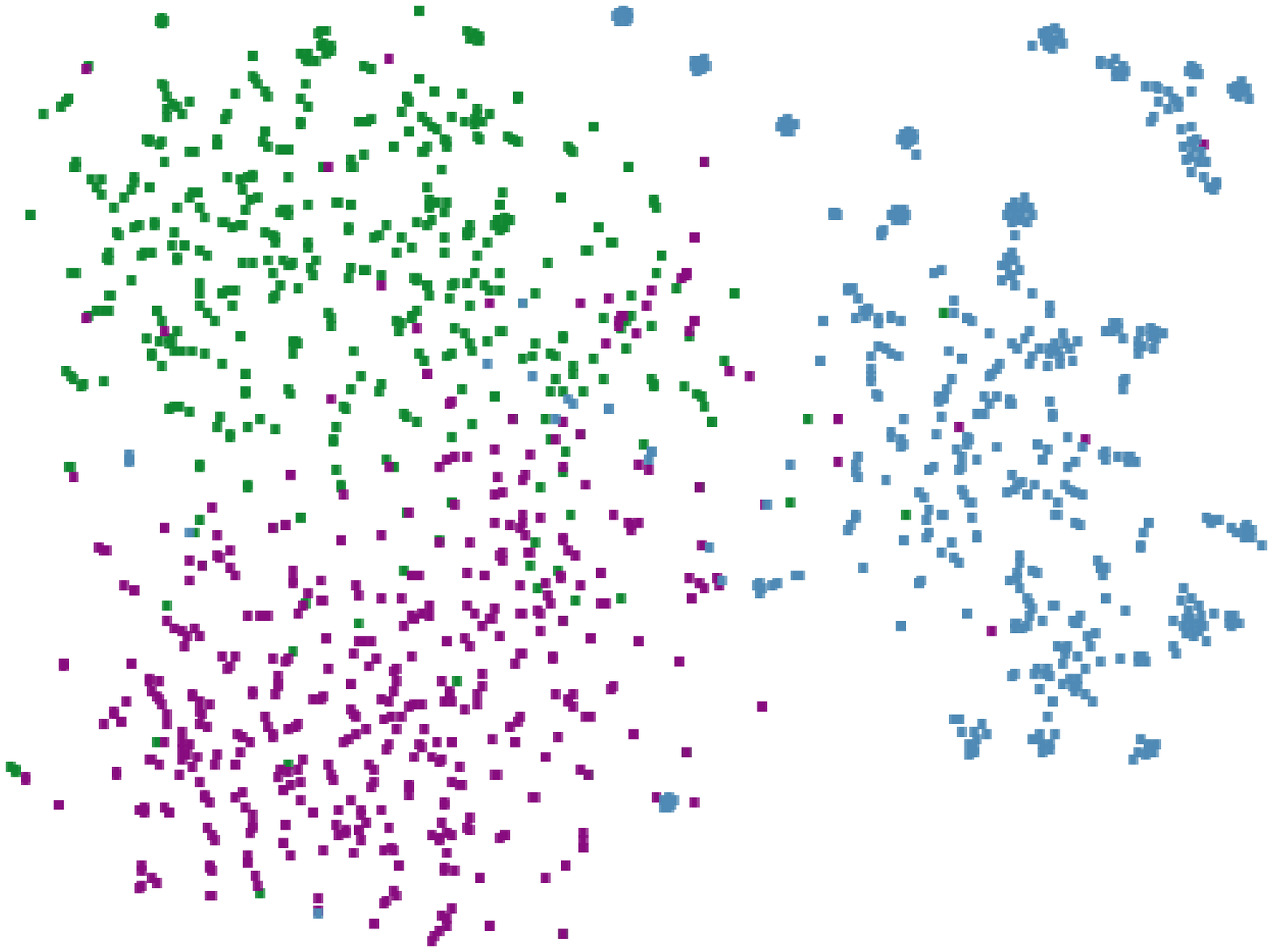}
\centerline{(b) node2vec}
\end{minipage}%
\begin{minipage}[t]{0.16\textwidth}
\includegraphics[width=1\textwidth]{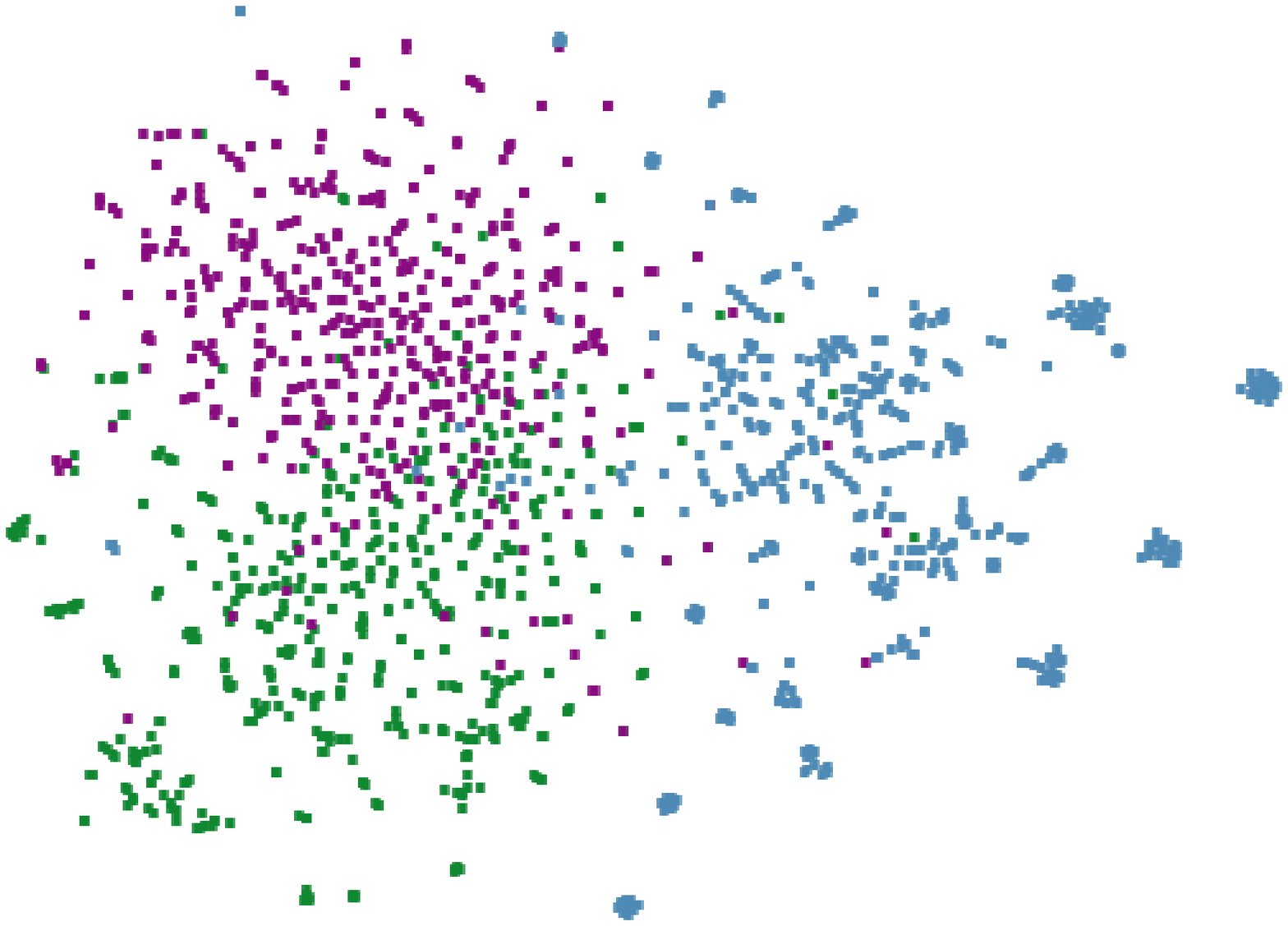}
\centerline{(c) GraphSAGE}
\end{minipage}%
~\\~\\~\\
\begin{minipage}[t]{0.16\textwidth}
\includegraphics[width=1\textwidth]{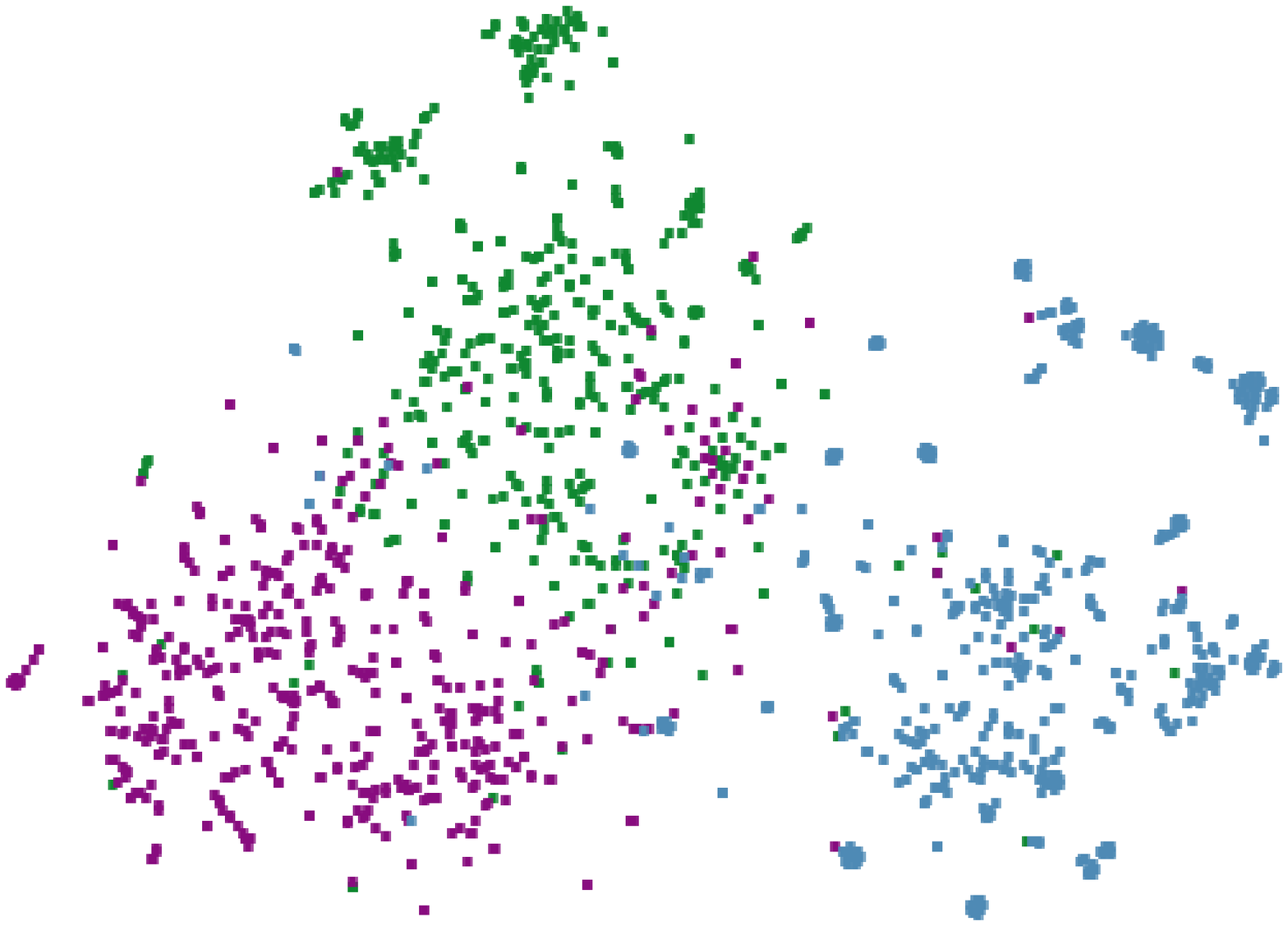}
\centerline{(d) HTNE}
\end{minipage}%
\begin{minipage}[t]{0.16\textwidth}
\includegraphics[width=1\textwidth]{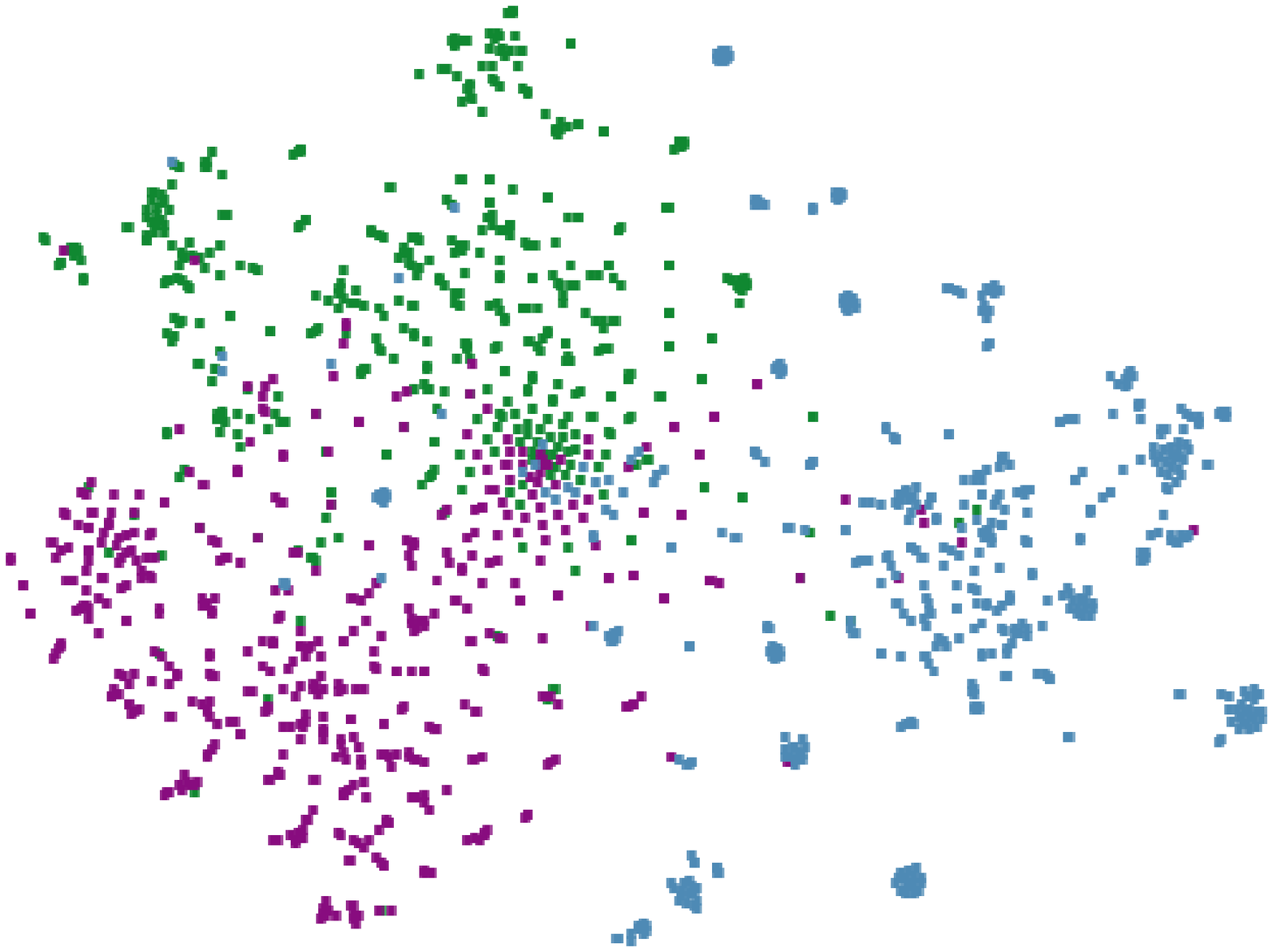}
\centerline{(e) DyREP}
\end{minipage}%
\begin{minipage}[t]{0.16\textwidth}
\includegraphics[width=1\textwidth]{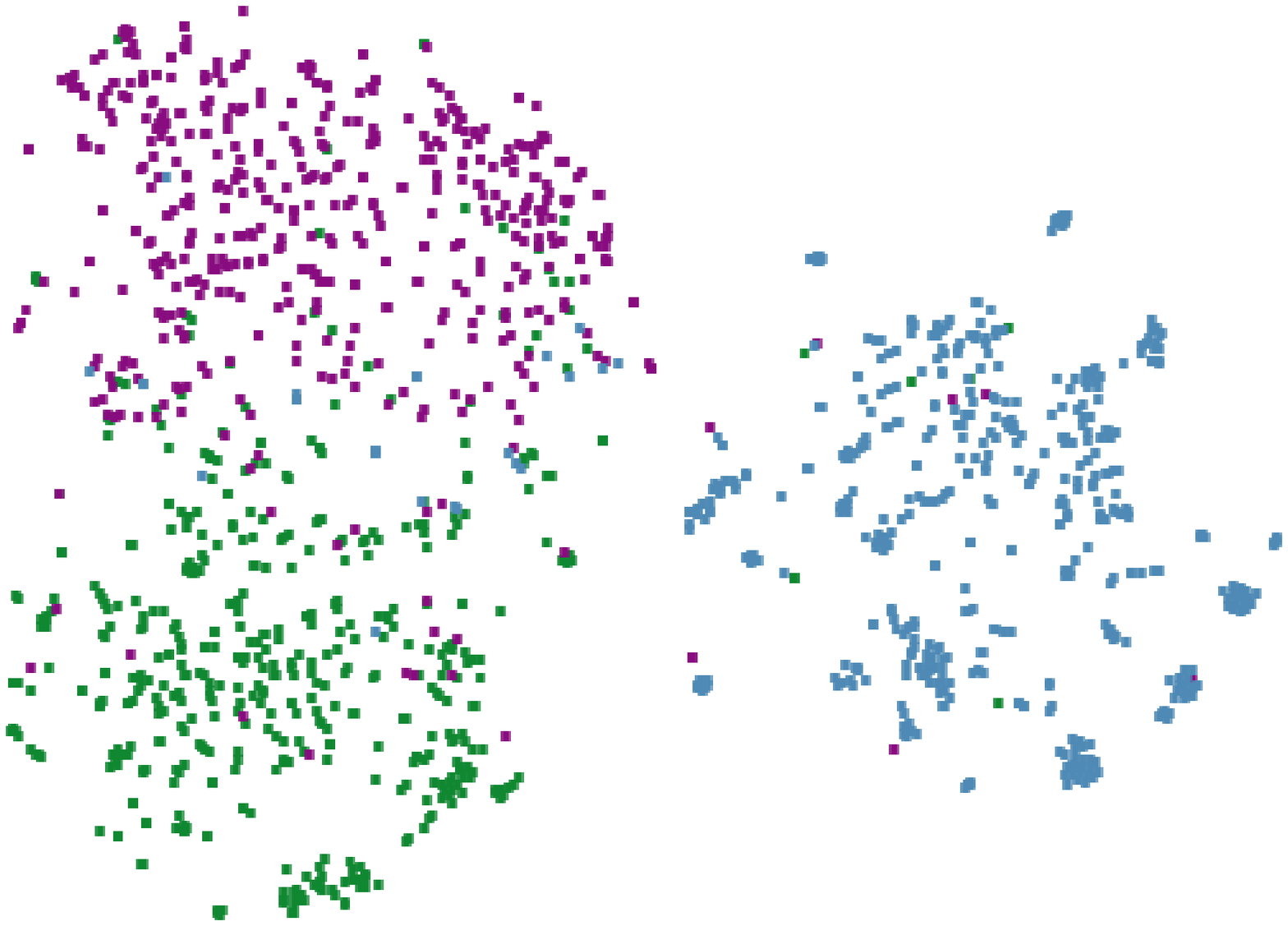}
\centerline{(f) MNCI}
\end{minipage}%
\caption{Network visualization}
\label{network}
\end{figure}

\subsubsection{Network visualization} 

We employ the t-SNE method \cite{maaten2008visualizing} to project node embeddings on DBLP to a 2-dimensional space. In particular, we select three fields and 500 researchers in each field. Selected researchers are shown in a scatter plot, in which different fields are marked with different colors, i.e., green for data mining, purple for computer vision, blue for computer network. As shown in Figure \ref{network}, both DeepWalk, node2vec, and GraphSAGE failed to separate the three fields clearly. HTNE and DyREP can only roughly distinguish the field boundaries. MNCI separates the three fields clearly, and one of them has a clear border, which indicates that MNCI has better performance.

\section{Conclusions}
We propose an inductive network representation learning method MNCI that captures both neighborhood and community influences to generate node embeddings at any time. Extensive experiments on several real-world datasets demonstrate that MNCI significantly outperforms state-of-the-art baselines. In the future, we will study the influence of node text information on node embeddings.

\section{Acknowledgment}
This work was supported by the National Natural Science Foundation of China (No. 61972135), 
the Natural Science Foundation of Heilongjiang Province in China (No. LH2020F043), 
and the Innovation Talents Project of Science and Technology Bureau of Harbin (No. 2017RAQXJ094).

\bibliographystyle{ACM-Reference-Format}
\bibliography{SIGIRshort}

\end{document}